\documentclass[10pt]{article}
\usepackage{amsmath}
\usepackage{graphicx}
\usepackage{amsfonts}
\usepackage{amssymb}
\usepackage{epsf}
\usepackage{latexsym}

\def\80{\hspace{0.8in}}

\newcommand{\be}{\begin{enumerate}}
\newcommand{\ee}{\end{enumerate}}
\newcommand{\bi}{\begin{itemize}}
\newcommand{\ei}{\end{itemize}}
\newcommand{\bd}{\begin{description}}
\newcommand{\ed}{\end{description}}

\def\beq{\begin{equation}}
\def\eeq{\end{equation}}
\def\bea{\begin{eqnarray}}
\def\eea{\end{eqnarray}}
\def\hat{\widehat}
\def\tilde{\widetilde}

\def\pa{\partial}
\def\d{\textrm{d}}

\def\tea{\mbox{\tt t}}
\def\stea{\mbox{\scriptsize\tt t}}

\def\ttQ{\mbox{\tt Q}}

\def\ttL{\mbox{\tt L}}
\def\ttD{\mbox{\tt D}}
\def\ttP{\mbox{\tt P}}

\def\Circ{\mbox{\Large$\circ$}}
\def\ncirc{\mbox{$\circ$}}
\def\scirc{\mbox{\scriptsize$\circ$}}
\def\Star{\mbox{\Large$\ast$}}

\def\cr{\mbox{\scriptsize{\bf $\mbox{ } \times \mbox{ }$}}}

\def\mF{\mbox{F}}

\def\mI{\mbox{I}}

\def\sa{\mbox{\scriptsize a}}
\def\sb{\mbox{\scriptsize b}}
\def\sc{\mbox{\scriptsize c}}
\def\sd{\mbox{\scriptsize d}}
\def\se{\mbox{\scriptsize e}}
\def\sf{\mbox{\scriptsize f}}
\def\sg{\mbox{\scriptsize g}} 
 
\def\si{\mbox{\scriptsize i}}

\def\sll{\mbox{\scriptsize l}}  
\def\sn{\mbox{\scriptsize n}} 
\def\so{\mbox{\scriptsize o}}

\def\sr{\mbox{\scriptsize r}}

\def\st{\mbox{\scriptsize t}}

\def\sw{\mbox{\scriptsize w}}

\def\sA{\mbox{\scriptsize A}} 
\def\sB{\mbox{\scriptsize B}}
\def\sC{\mbox{\scriptsize C}}
\def\sD{\mbox{\scriptsize D}}

\def\sF{\mbox{\scriptsize F}}
\def\sG{\mbox{\scriptsize G}}

\def\sI{\mbox{\scriptsize I}}
\def\sJ{\mbox{\scriptsize J}}

\def\sL{\mbox{\scriptsize L}} 
\def\sM{\mbox{\scriptsize M}} 
\def\sN{\mbox{\scriptsize N}} 
\def\sO{\mbox{\scriptsize O}}

\def\sR{\mbox{\scriptsize R}}
\def\sS{\mbox{\scriptsize S}}

\def\sW{\mbox{\scriptsize W}}

\def\eph(B){\mbox{\scriptsize emergent(LMB)}}

\def\fA{\mbox{\sffamily A}}

\def\fE{\mbox{\sffamily E}}

\def\fH{\mbox{\sffamily H}}
\def\fI{\mbox{\sffamily I}}

\def\fQ{\mbox{\sffamily Q}}
\def\fR{\mbox{\sffamily R}}

\def\fT{\mbox{\sffamily T}}
\def\fU{\mbox{\sffamily U}}
\def\fV{\mbox{\sffamily V}}
\def\fW{\mbox{\sffamily W}}

\def\sfT{\mbox{\sffamily{\scriptsize T}}}

\def\sfW{\mbox{\sffamily{\scriptsize W}}}

\def\p{\underline{p}}
\def\q{\underline{q}}

\begin{document}
\begin{titlepage}
\vspace{.7in}
\begin{center}

\LARGE{\bf RELATIONAL MOTIVATION FOR CONFORMAL OPERATOR ORDERING IN QUANTUM COSMOLOGY} 

\vspace{.4in}

\large{\bf Edward Anderson}$^{1}$

\vspace{.2in}

\large{\em DAMTP, Centre for Mathematical Sciences, Wilberforce Road, Cambridge CB3 OWA.}

\end{center}

\begin{abstract}

Operator-ordering in quantum cosmology is a major as-yet unsettled ambiguity with not only formal but 
also physical consequences.  
We determine the Lagrangian origin of the conformal invariance that underlies the conformal 
operator-ordering choice in quantum cosmology.  
It is particularly naturally and simply manifest in relationalist product-type actions (such as the 
Jacobi action for mechanics or Baierlein--Sharp--Wheeler type actions for general relativity), for 
which all that is required for the kinetic and potential factors to rescale in compensation to each 
other.
These actions themselves mathematically sharply implementing philosophical principles relevant to 
whole-universe modelling, the motivation for conformal operator-ordering in quantum cosmology is 
substantially strengthened.
Relationalist product-type actions also give emergent times which amount to recovering Newtonian, proper 
and cosmic time in the various relevant contexts.  
The conformal scaling of these actions directly tells us how emergent time scales; if one follows suit 
with the Newtonian time or the lapse in the more commonly used difference-type Euler--Lagrange or 
Arnowitt--Deser--Misner type actions, one sees how these too obey a more complicated conformal 
invariance.  
Moreover, our discovery of the conformal scaling of the time involved permits relating how it simplifies 
equations of motion with how affine parametrization simplifies geodesics.

\end{abstract}

\vspace{1in}


\mbox{ }

\vspace{2in}

\noindent$^1$ ea212@cam.ac.uk

\end{titlepage}

%========================================================================================================
%========================================================================================================
\section{Introduction}
%========================================================================================================
%========================================================================================================

%========================================================================================================
\subsection{Relationalism and Product-type actions.} 
%========================================================================================================

Classical physics is usually taken to follow from difference-type actions\footnote{$\Sigma$ 
%%%%%%%%%%%%%%%%%%%%%%%%%%%%%%%%%%%%%%%%%%%%%%%%%%%%%%%%%%%%%%%%%%%%%%%%%%%%%%%%%%%%%%%%%%%%%%%%%%%%%%%%%
is the notion of space of extent (3-space for field theories and trivial for particle theories, for 
which $\int_{\Sigma}\d\Sigma$ is taken to become $\times 1$).  
$Q^{\Lambda}$ are generalized coordinates with $\Gamma$ a multi-index over particle and/or field species 
as well as over space of extent.  
I use round brackets for functions and square brackets for functionals.} 
%%%%%%%%%%%%%%%%%%%%%%%%%%%%%%%%%%%%%%%%%%%%%%%%%%%%%%%%%%%%%%%%%%%%%%%%%%%%%%%%%%%%%%%%%%%%%%%%%%%%%%%%% 
\beq
\fI = \int\d \tea\int_{\Sigma}\d\Sigma\{\fT_{\stea} - \fV\}
\mbox{ } .
\label{diff}
\eeq
Here, $\fT_{\stea}$ is the kinetic terms that is usually\footnote{One can include a sufficient set of 
%%%%%%%%%%%%%%%%%%%%%%%%%%%%%%%%%%%%%%%%%%%%%%%%%%%%%%%%%%%%%%%%%%%%%%%%%%%%%%%%%%%%%%%%%%%%%%%%%%%%%%%%%
fields in this context to study classical fundamental physics (there being no difficulty 
\cite{Van, Phan} with additionally incorporating terms linear in the velocities into this scheme and 
this paper's workings, e.g. permitting fermionic as well as bosonic matter coupled to GR.}
%%%%%%%%%%%%%%%%%%%%%%%%%%%%%%%%%%%%%%%%%%%%%%%%%%%%%%%%%%%%%%%%%%%%%%%%%%%%%%%%%%%%%%%%%%%%%%%%%%%%%%%%% 
homogeneous quadratic in the velocities,  
\beq
\fT_{\stea} = \frac{1}{2}\sum \mbox{}_{\mbox{}_{\Gamma, \Delta}}M_{\Gamma\Delta}\frac{\d Q^{\Gamma}}{\d \tea}
                                                                    \frac{\d Q^{\Delta}}{\d \tea}
\mbox{ } , 
\eeq
where $M_{\Gamma\Delta} = M_{\Gamma\Delta}(Q^{\Lambda})$ is the kinetic metric of the configuration 
space and  $\fV[Q^{\Lambda}]$ is the potential term.

However, classical physics can also be taken to follow from product-type actions
\beq
\fI = 2\int\d \lambda\int_{\Sigma}\d \Sigma\sqrt{\fT\fW} \mbox{ } .   
\label{prod}
\eeq
Here, $\fT$ takes the form 
\beq
\fT = \sum \mbox{}_{\mbox{}_{\Gamma, \Delta}}
M_{\Gamma\Delta}\Circ Q^{\Gamma}\Circ Q^{\Delta}/2 \mbox{ } , 
\eeq 
$\Circ = \d/\d\lambda$ and $\lambda$ is a label-time, and $\fW[Q^{\Lambda}]$ is minus the potential 
(possibly up to an additive constant energy, as explained in the examples below).  
Action (\ref{diff}) leads to action (\ref{prod}) by, firstly (parametrization): adjoining $\tea$ to the 
configuration space so that $\d \tea/\d\lambda$ now features in it. 
Secondly, provided that $\fV$ is independent of $\tea$ and $\d Q^{\Gamma}/\d \tea$, which can be held to be the 
case when one is considering fundamental classical physics of the universe as a whole, Routhian 
reduction \cite{Lanczos} subsequently serves to eliminate $\d \tea/\d\lambda$ from the variational equation 
for $\tea$.

Product-type actions have advantages over difference-type actions for consideration of whole-universe 
fundamental physics -- the setting for quantum cosmology.  
Product-type actions arise if one sets up physics from the Leibniz--Mach--Barbour relational first 
principles, which both have philosophical significance and are sharply mathematically implementable.   
Relationalism provides an alternative foundation for physics to absolutism; which of these to use has 
been the subject of a long debate \cite{Newton, AORM}, and relationalism would appear to have a good 
case for use in whole-universe situations.  
I use relationalism in this Leibniz--Mach--Barbour's sense of the word \cite{AORM, BB82, B94I, EOT, 
RWR, fqxi, Ak}; see \cite{Rovelli} for Rovelli's distinct use of the same word and \cite{08I} for a 
brief comparison of the two.  
The Leibniz--Mach--Barbour relational first principles are as follows.  

\noindent
A physical theory is {\it temporally relational} if there is no meaningful primary notion of time for 
the whole system thereby described (e.g. the universe) \cite{BB82, RWR}.  
This is implemented by using actions that are {\it manifestly reparametrization invariant} while also 
being free of extraneous time-related variables (such as external Newtonian time or the 
geometrodynamical formulation of general relativity (GR)'s lapse coordinate \cite{MTW}).   

\noindent 
A physical theory is {\it configurationally relational} if a certain group $G$ of transformations that 
act on the theory's configuration space $\fQ$ are physically meaningless \cite{BB82, RWR, 
ABABFOLanThanABFKO, Van, B03, FORD}.  
As subcases, this includes {\it spatially relational} and {\it internally relational} (in the usual 
gauge-theoretic sense) depending on the mathematical form and physical interpretation of $G$. 
Spatial relationalism suffices for the examples covered in this paper.

It is temporal relationalism that is directly tied to the product-type actions in this paper; the 
reparametrization invariance of these actions is clear since changing from $\lambda$ to another 
$\lambda^{\prime}$ clearly cancels as $\fT$ is homogeneous quadratic.  
Indeed, one could implement temporal relationalism, rather, in a {\it parametrization irrelevant} way
(i.e. one which makes no reference whatsoever to any label-time parameter $\lambda$):
\beq
\fI = 2\int\int_{\Sigma}\d \Sigma\sqrt{\fW\,\sum \mbox{}_{\mbox{}_{\Gamma, \Delta}}
M_{\Gamma\Delta}\d Q^{\Gamma}\d Q^{\Delta}/2} \mbox{ } .   
\eeq

Starting from product-type actions,  using a $\tea^{\sJ\sB\sB}$ 
(`Jacobi--Barbour--Bertotti time') \cite{B94I, SemiclI, SemiclII, fqxi} such that 
\beq
{\d}/{\d \tea^{\sJ\sB\sB}} = \sqrt{\fW/\fT}\Circ = 
\sqrt{\fW}\d/\sqrt{\sum \mbox{}_{\mbox{}_{\Gamma,\Delta}}M_{\Gamma\Delta}\d Q^{\Gamma}\d Q^{\Delta}/2}
\eeq
{\sl is found} to considerably simplify the equations of motion that follow from the relational principles.  
In this sense such a $\tea^{\sJ\sB\sB}$ is a privileged parametrization; it {\sl coincides} with the 
conventional difference action's $\tea$, but is now to be considered as {\sl emergent and provided by the 
entirety of the model universe's contents}.\footnote{Barbour
%======================================================================================================== 
furthermore considers \cite{B94I, fqxi} this to be the timestandard such that isolated observers who 
choose to use it obtain clocks that march in step with each others'.  
However, as I have never seen this quantitativly demonstrated, it will play no further part in this 
paper.}
%========================================================================================================

A further point of note is that product-type actions' reparametrization invariance gives as a primary 
constraint  
\beq
N^{\Gamma\Delta}P_{\Gamma}P_{\Delta}/2 + \fV = \fE \mbox{ } ,   
\label{set}
\eeq
which is quadratic in the momenta. [Here $N^{\Gamma\Delta}$ is the inverse of $M_{\Gamma\Delta}$.]

%========================================================================================================
\subsection{Examples of product-type actions}
%========================================================================================================

\noindent{\bf Example 1)} The Jacobi action \cite{Lanczos} for the Newtonian mechanics of N particles 
with positions $\underline{q}_I$, $I$ = 1 to N is
\beq
\fI_{\sJ\sa\sc\so\sb\si} = 2\int\d\lambda\sqrt{\fT\{\fU + \fE\}} \mbox{ } . 
\label{7}
\eeq
Here 
\beq
\fT = \frac{1}{2}\sum\mbox{}_{\mbox{}_{I = 1}}^{\sN}m_I\{\Circ q_I\}^2 \mbox{ } , 
\eeq
$\fU(Q_{\Lambda})$ is minus the potential term, $\fV$, and $\fE$ is the total energy of the system.  
\cite{Lanczos} lucidly covers how to obtain this from $\fI = \int\d t\{\fT_t - \fV\}$ by Routhian 
reduction.   
(I use $t$ in place of $\tea$ for mechanical theories).  
For the recovery from (\ref{7}) of the usual Euler--Lagrange formalism from this but with the emergent 
$t^{\sJ\sB\sB}$ now taking over the role of Newtonian absolute time, see e.g. \cite{B94I, SemiclI}.  
The quadratic constraint is in this case the energy constraint $ \sum_{I = 1}^Np_I^2/2 m_I+ \fV = \fE$.  

%\mbox{ }

\noindent{\bf Example 2A)} One could consider versions of the Jacobi action for `relational particle 
mechanics' theories in which the velocities come with arbitrary Euclidean \cite{BB82, B94I, EOT, GGM06I, 
ParisTriCl, 08I, Cones, scaleQM, 08III, Ultra} or similarity \cite{B03, Piombino, ParisTriCl, 06II, 
FORD, 08I, 08II, AF, +tri, Ultra} group frame corrections\footnote{Here,
%%%%%%%%%%%%%%%%%%%%%%%%%%%%%%%%%%%%%%%%%%%%%%%%%%%%%%%%%%%%%%%%%%%%%%%%%%%%%%%%%%%%%%%%%%%%%%%%%%%%%%%%%
$\stackrel{\longrightarrow}{G_{\scirc{g}}}$ is the group action corresponding to an infinitesimal group 
generator $\ncirc{g}$; note that this notation extends from the current relational particle 
mechanics context of the current example 
to the GR case of Example 3 as well.}
%%%%%%%%%%%%%%%%%%%%%%%%%%%%%%%%%%%%%%%%%%%%%%%%%%%%%%%%%%%%%%%%%%%%%%%%%%%%%%%%%%%%%%%%%%%%%%%%%%%%%%%%% 
\beq
\Circ_gQ^{\Delta} \equiv {\d_g Q^{\Delta}}/{\d \lambda} \equiv 
\Circ{Q}^{\Delta} - \stackrel{\longrightarrow}{G_{\ncirc{g}}}Q^{\Delta}
\eeq
that implement spatial relationalism as well, so that one is considering mechanical theories that are 
temporally {\sl and} spatially relational.  
The quadratic constraint continues in this case to be an energy constraint of the form in the preceding 
example.  
Now one is to use $\d_g/\d t$ in place of $\d/\d t$ in (2), $\Circ_g$ in place of $\Circ$ in (4, 8) and $\d_g Q^{\Gamma}$ in place of 
$\d Q^{\Gamma}$ in (5) and (6).  
Also, variation with respect to the auxiliary variables $g$ produces constraints that are linear in the 
momenta.  
For Euclidean relational particle mechanics, these are 
$\underline{\ttP} = \sum_{I = 1}^{\sN}\p_I = 0$ (zero total momentum for the model universe) and 
$\underline{\ttL} = \sum_{I = 1}^{\sN}\q_I \cr \p_I = 0$ (zero total momentum for the model universe), 
while similarity relational particle mechanics has these again alongside $\ttD = \sum_{I = 1}^{\sN}
\q_I\cdot \p_I = 0$.  
By analogy with Example 3 below, relational particle mechanics are additionally useful models \cite{BS89, K92, B94II, Kieferbook, 
06II, SemiclI, SemiclII, Records, SemiclIII, Smolin08, BF08Gryb, 08II, 08III} for the Problem of Time 
in Quantum Gravity \cite{K92, I93} and other issues of interest in Quantum Cosmology \cite{DeWitt67, 
QCosLit, Wiltshire}.  

%\mbox{ }

\noindent{\bf Example 2B)} Relational particle mechanics can be cast in reduced form in spatial 
dimension 1 or 2.  
Here all the above constraints can be eliminated, producing reduced kinetic terms of the form 
$$
\fT^{\sr\se\sd} = M_{\Gamma\Delta}(Q^{\Lambda})\Circ{Q}^{\Gamma}\Circ{Q}^{\Delta}/2
$$
for $M_{\Gamma\Delta}$ the usual metric on $\mathbb{S}^{\sN - 2}$ for N particles in 1-d and the 
Fubini--Study metric on $\mathbb{CP}^{\sN - 2}$ for N particles in 2-d.  
The quadratic constraint is still an energy constraint of form (\ref{set}) built from inverses of the 
above-mentioned metrics.  

%\mbox{ }

\noindent{\bf Example 3A)} An action (see e.g. \cite{RWR, Phan}) for a geometrodynamical formulation of 
GR [in terms of 3-metrics $h_{\mu\nu}(x^{\omega})$ on a fixed topology $\Sigma$, for simplicity taken to be compact without 
boundary; $x^{\omega}$ are spatial coordinates] is  
\beq
\fI^{\sB\sF\sO-\sA}_{\sG\sR} = 2\int\d\lambda\int_{\Sigma}\d^3{x}\sqrt{h}\sqrt{\fT_{\sB\sF\sO-\sA}
\{\mbox{Ric}(h) - 2\Lambda\}} \mbox{ } .
\label{BFOA}
\eeq
Here, 
\beq
\fT^{\sB\sF\sO-\sA}_{\sG\sR} = 
{\cal M}^{\mu\nu\rho\sigma}\Circ_{\sF}h_{\mu\nu}\Circ_{\sF}h_{\rho\sigma}/4 
\mbox{ } , \mbox{ } 
\Circ_{\sF}h_{\mu\nu} \equiv \Circ{h}_{\mu\nu} - 
\stackrel{\longrightarrow}{\mbox{Diff}_{\ncirc{\sF}}}h_{\mu\nu} = 
\Circ{h}_{\mu\nu} - \pounds_{\ncirc{\sF}}h_{\mu\nu} \mbox{ } ,
\label{TBFO}
\eeq 
where ${\cal M}^{\mu\nu\rho\sigma} = h^{\mu\rho}h^{\nu\sigma} - h^{\mu\nu}h^{\rho\sigma}$ (the GR 
configuration space metric, alias inverse of the undensitized DeWitt supermetric \cite{DeWitt67}), Diff 
is the group of 3-diffeomorphisms on $\Sigma$, $\pounds_{\ncirc{\sF}}$ is the Lie derivative with respect 
to the `velocity of the frame' $\mF_{\mu}$, 
Ric($h$) is the Ricci 3-scalar corresponding to $h_{\mu\nu}$, $h$ is the determinant of $h_{\mu\nu}$ and 
$\Lambda$ is the cosmological constant.  
This action would be the better-known Baierlein--Sharp--Wheeler (BSW) \cite{BSW} one if the kinetic 
term were, rather, $\fT_{\sB\sS\sW}$ which is the same up to being built out of shift corrections 
$\beta^{\mu}(x^{\omega})$ in place of `velocities of the frame' 
$\Circ{\mF}^{\mu}(x^{\omega})$.\footnote{There are various
%%%%%%%%%%%%%%%%%%%%%%%%%%%%%%%%%%%%%%%%%%%%%%%%%%%%%%%%%%%%%%%%%%%%%%%%%%%%%%%%%%%%%%%%%%%%%%%%%%%%%%%% 
other equivalent pairs of Principles of Dynamics objects in this paper that are related to each other 
by the one using auxiliary frame velocities where the other uses auxiliary coordinates (and, sometimes 
additionally, auxiliary {\sl instant} velocities, $\ncirc{\sI}$, in place of of auxiliary lapse coordinates, 
$\alpha$).  
For details of how the equivalence of each of these pairs works out, see \cite{FEPI}.}    
%%%%%%%%%%%%%%%%%%%%%%%%%%%%%%%%%%%%%%%%%%%%%%%%%%%%%%%%%%%%%%%%%%%%%%%%%%%%%%%%%%%%%%%%%%%%%%%%%%%%%%%%
However it would not then be manifestly temporally relational.    
Moreover, the BSW action is equivalent to the even more familiar `Lagrangian ADM' \cite{ADM} action, 
\beq
\fI^{\sL-\sA\sD\sM}_{\sG\sR} = \int\d\lambda\int_{\Sigma}\d^3x\sqrt{h}\alpha 
\{{\fT_{\sA\sD\sM}}/{\alpha^2} + \mbox{Ric}(h) - 2\Lambda\}  
\label{ADM}
\eeq  
(for $\fT_{\sA\sD\sM}$ taking the same form as the above $\fT_{\sB\sS\sW}$).  
The former follows from the latter by elimination of the Lagrange multiplier coordinate lapse $\alpha$ 
from its own variational equation.  
Parallely \cite{FEPI}, one can also obtain the BFO-A action from a now also-unfamiliar action that is the 
Lagrange--ADM's equivalent pair in the sense of footnote 4, 
\beq
\fI^{\sA}_{\sG\sR} = \int\d\lambda\int_{\Sigma}\d^3x\sqrt{h}\Circ{\mI} 
\{ \fT^{\sA}_{\sG\sR}/\{\Circ{\mI}\}^2 + \mbox{Ric}(h) - 2\Lambda\}    
\label{A}
\eeq  
(for $\fT^{\sA}_{\sG\sR}$ taking the same form as $\fT_{\sB\sF\sO-\sA}$).  
The former now follows from the latter by using Routhian reduction to eliminate $\dot{I}$ (an even 
closer parallel of the equivalence at the end of Example 1 than the preceding coordinate elimination).   
The quadratic constraint is now the GR Hamiltonian constraint ${\cal H} \equiv 
{\cal N}_{\mu\nu\rho\sigma}\pi^{\mu}\pi^{\nu}/\sqrt{h} - \sqrt{h}\{\mbox{Ric}({\cal M}) - 2\Lambda\} 
= 0$ for ${\cal N}^{\mu\nu\rho\sigma}$ the inverse of ${\cal M}_{\mu\nu\rho\sigma}$ (i.e. the 
undensitized DeWitt supermetric itself), while the linear constraint from variation with respect to 
$F^{\mu}$ is the GR momentum constraint, ${\cal L}_{\mu} \equiv - 2D_{\nu}\pi^{\nu}\mbox{}_{\mu} = 0$.  

%\mbox{ }

{\noindent \bf Example 3B.}  
Each of the pairs (\ref{BFOA}, BSW) and (\ref{ADM}, \ref{A}), A become indistinguishable for minisuperspace. 
Therein, ${\cal M}^{\mu\nu\rho\sigma}(h_{\gamma\delta}(x^{\omega}))$ collapses to an ordinary 
$6 \times 6$ matrix $M_{\Gamma\Delta}$ or further in the diagonal case (a $3 \times 3$ matrix 
$M_{\Gamma\Delta}$) -- the `minisupermetric'.  
%

%========================================================================================================
\subsection{Banal-conformal invariance of product-type actions and its consequences.}
%========================================================================================================

In Sec 2, I start at the classical level, both to set up the foundations for the paper's principal 
quantum cosmology operator ordering issue and also to consider a second classical issue concerning 
parametrization of dynamical curves.  
In considering these two applications together, I follow Misner's Hamiltonian treatment \cite{Magic}; 
my work differs from his in considering the Lagrangian formulation and the Jacobi formulation, which 
sits on relational first principles.

I show that product-type actions are preserved under the simple and natural 
{\it banal conformal transformation}  
\beq
\fT \longrightarrow \widetilde{\fT} = \Omega^2 \fT \mbox{ } 
\mbox{ } , \mbox{ } \mbox{ } 
\fW \longrightarrow \widetilde{\fW} = \fW/\Omega^2 \mbox{ }  .  
\eeq
This makes the kinetic factor a banal-vector and the potential factor a banal covector.  
Moreover the first of these can be viewed as 
\beq
M_{\Gamma\Delta} \longrightarrow \widetilde{M}_{\Gamma\Delta} = \Omega^2M_{\Gamma\Delta} \mbox{ } ,
\eeq
so that the kinetic metric is a banal-vector.
It immediately follows from the above banal transformation that the emergent timefunction can be 
regarded as a banal covector.  
If one then considers this scaling property to carry over to the difference-type action formulations' 
timefunction, a more complicated manifestation of banal-conformal invariance is discovered for 
difference-type actions.
Clearly, performing such a transformation should not (and does not) affect one's classical equations of 
motion.
Moreover, working through how the scaling of $\fT$, $\fW$ and the timefunction conspire to cancel out 
at the level of the classical equations of motion reveals interesting connections between the 
simplifying effects of using the emergent timefunction on the equations of motion and those of the 
rather better-known affine parametrization \cite{Wald, Stewart}.  
Section 2 ends by preparing for quantization by discussing how momenta, constraints and 
Hamiltonian-type objects banal-scale.

If one then (Sec 3) wishes for this banal-conformal invariance -- displayed simply and naturally by 
relationalism-implementing product actions for whole-universe fundamental physics -- to continue to 
hold at the {\sl quantum} level, then this alongside the otherwise theoretically-desirable (and fairly 
standard) requirement that one's quantum theory should not depend on how $\fQ$ is coordinatized, then 
one is led to the operator ordering for $N^{\Gamma\Delta}(Q^{\Lambda})P_{\Gamma}P_{\Delta}$ that is based 
on the conformally-invariant modification of the Laplacian.     
The latter requirement is due to DeWitt \cite{DeWitt57} (see also \cite{Magic, K73, HPT, HP86}), and is 
true for 1-parameter family of scalar operators 
\beq
\nabla^2 - \xi\mbox{Ric($M$)}
\label{Family}
\eeq
where $\mbox{Ric($M$)}$ is the Ricci scalar corresponding to the configuration space metric 
$M_{\Gamma\Delta}$ and $\nabla^2$ is the Laplacian,  
\beq
\nabla^2 = \frac{1}{\sqrt{M}} \frac{\nabla}{\nabla {Q}^{\Gamma}}
\left\{
\sqrt{M}N^{\Gamma\Delta}\frac{\nabla}{\nabla{Q}^{\Delta}}
\right\} \mbox{ } . 
\eeq  
Here, $\nabla$ denotes partial derivative for finite theories and functional derivative for field 
theories and $\sqrt{M} = \sqrt{\mbox{det}(M)}$.  
The conformal ordering, which fixes a particular value of $\xi$, had been previously suggested by e.g. 
Misner \cite{Magic}, Halliwell \cite{Halliwell}, Moss \cite{Moss} and Ryan--Turbiner \cite{RT}, 
albeit without any reference to the immediacy of this in product-type actions which themselves rest 
on the relationalist first principles.   
Additionally, Kucha\v{r} \cite{K73} and Henneaux--Pilati--Teitelboim \cite{HPT} have advocated the Laplacian 
ordering itself ($\xi = 0$).  
So have Page \cite{Page91}, Louko \cite{Louko} and Barvinsky \cite{Barvin}, however their specific 
examples are 2-dimensional, for which the Laplacian and conformal orderings coincide. 
Wiltshire advocates both \cite{Wiltshire}. 
Christodoulakis and Zanelli \cite{CZ} consider the case with an arbitrary $\xi$, as do Hawking and 
Page \cite{HP86}, albeit the latter then also pass to a 2-d example for which $\xi$ drops out.  
Finally, all of these orderings coincide to O($\hbar$) \cite{Barvin}.

The reason why arguments for such a choice of operator ordering is of interest is the well-known 
physical as well as just formal inequivalence of different operator orderings, so that how to make 
such a choice is a major issue in Quantum Gravity and Quantum Cosmology \cite{DeWitt57, DeWitt67, 
Magic, K73, HPT, HP86, CZ, Halliwell, Moss, Page91, Louko, Barvin, K92, I93, RT, K78Thiemann}. 
Thus my answer that relationalism and coordinate invariance combine to imply the elegant and 
mathematically well-distinguished conformal ordering should be of considerable interest.
In being revealed to possess philosophical foundations of this kind alongside the technical advantages 
already found in the above-cited papers, I argue that the case to adopt the conformal ordering is 
considerably strengthened.

%========================================================================================================
%========================================================================================================
\section{Banal invariance of product-type actions, and classical consequences}
%========================================================================================================
%========================================================================================================

The equations of motion that follow from the spatially relational generalization of (3) are the 
momentum--velocity relations
\beq
P_{\Gamma} = \sqrt{{\fW}/{\fT}}M_{\Gamma\Delta}\Circ_g Q^{\Delta}
\eeq
and the Euler--Lagrange equations 
\beq
\Circ
\left\{
\sqrt{{\fW}/{\fT}}\,\Circ_gQ^{\Gamma}
\right\} + \sqrt{\fW/\fT}\,{\Gamma^{\Gamma}}\mbox{}_{\Delta\Lambda}\Circ_gQ^{\Delta}\Circ_gQ^{\Lambda} = 
\sqrt{{\fT}/{\fW}}\,\nabla^{\Gamma}\fW + M_{\Delta\Lambda}\sqrt{\fW/\fT}\,\Circ_{g}Q^{\Delta}
\nabla^{\Gamma}\stackrel{\longrightarrow}{\{G}_{\Circ{g}}Q_{\Lambda}\} 
\label{tw}
\eeq
(for $\nabla^{\Gamma} \equiv \nabla/\nabla Q_{\Gamma}$ and $\Gamma^{\Gamma}\mbox{}_{\Delta\Lambda}$ the 
configuration space Christoffel symbols).

Upon inspection (see e.g. \cite{B94I}), (\ref{tw}) simplifies for particular choices of parameter 
in 2 generally different ways (the two coincide if the potential is constant).  

\noindent 
A)
$$
\frac{\d}{\d\lambda}
\left\{ 
\sqrt{\frac{\fW}{\fT}}\frac{\d Q^{\Delta}}{\d\lambda}
\right\}
= \frac{\d^2Q^{\Delta}}{\d \lambda} + 
\frac{1}{2\sqrt{\fW\fT}}\frac{\d\fW}{\d\lambda}\frac{\d Q^{\Delta}}{\d\lambda} - 
\frac{1}{2}\frac{\d \fT}{\d \lambda}\sqrt{\frac{\fW}{\fT^3}}\frac{\d Q^{\Delta}}{\d \lambda} 
\mbox{ versus } \frac{\d^2 Q^{\Gamma}}{\d \mu^2}
$$
which corresponds to $\frac{\d}{\d \mu} = \sqrt{\frac{\sfW}{\sfT}}\frac{\d}{\d\lambda}$, which parameter 
$\mu$ we denote by $\tea^{\sJ\sB\sB}$: emergent Jacobi--Barbour--Bertotti time \cite{B94I, RWR}.
In the case of a mechanical theory, this emergent time turns out 
also to imply conservation of energy and amounts to a recovery of Newtonian time; 
it is also aligned with the mechanics case's emergent semiclassical (WKB) time \cite{SemiclI}. 
In the case of geometrodynamics (I use $T$ in place of $\tea$ in this context), this emergent time amounts to 
a recovery of local proper time, as well as being aligned with the geometrodynamical emergent 
semiclassical (WKB) time \cite{SemiclII}, and corresponding to cosmic time in the case of homogeneous 
cosmology.

\noindent B) $\nabla^{\Gamma}\fW \neq 0$ versus = 0, the latter corresponding to `the dynamical curve 
being an affinely-parametrized geodesic on configuration space'.  
In this case I denote the time parameter by $\tea^{\sa\sf\sf-\sg\se\so}$.

Next, note the following properties.  

\mbox{ }

\noindent 1) The action (\ref{prod}) is banal-conformally invariant under 
\beq
\fT \longrightarrow \widetilde{\fT} = \Omega^2\fT \mbox{ } \mbox{ } , \mbox{ } \mbox{ }
\fW \longrightarrow \widetilde{\fW} = \fW/\Omega^2 \mbox{ } \mbox{ } :   
\eeq
$$
\widetilde{\fI} = 2\int\d\lambda\int_{\Sigma}\d\Sigma \sqrt{ \widetilde{\fT}\widetilde{\fW} } = 
2\int\d\lambda \int_{\Sigma}\d\Sigma \sqrt{ \Omega^2{\fT}\fW/\Omega^2} = 
2\int\d\lambda \int_{\Sigma}\d\Sigma \sqrt{\fT\fW} \mbox{ } .  
$$
2) Subsequently, ${\d}/{\d \mbox{(emergent time)}}$ is a banal-conformal covector, 
$$
\widetilde{\Star} \equiv \sqrt{\widetilde{\fW}/\widetilde{\fT}}\Circ = 
\sqrt{\{\fW/\Omega^2\}/{\Omega^2\fT}}\Circ = \Omega^{-2}\sqrt{{\fW}/{\fT}}\Circ = \Omega^{-2}\Star 
\mbox{ } .
$$
Thus the emergent time \cite{B94I, B94II, SemiclI} depends on the choice 
of banal-conformal factor. 
In the case of geometrodynamics, one can also think of ${\d}/{\d\lambda}$ being invariant and 
${1}/{\tilde{\Circ{I}}} = {1}/\{\Omega^2 \Circ{I}\}$ so that the velocity of the instant scales as a 
banal-conformal vector $\Circ{I} \longrightarrow \Omega^2\Circ{I}$ (or, equivalently, the emergent lapse 
coordinate scales as a banal-conformal vector $N \longrightarrow \Omega^2N$).

\noindent
To not confuse `$\tea^{\sJ\sB\sB}$ as present in the previous literature' and the banal covector 
discovered in this paper, I denote the latter by $\vec{\tea}$, rather.
I also use the notation
$$
\Star \equiv {\d}/{\d \vec{\tea}} = \sqrt{{\fW}/{\fT}}\Circ \mbox{ } . 
$$
 
\mbox{ }

\noindent 3) Next observe that, provided that its timefunction scales as $\vec{\tea}$ does, the 
difference-type Euler--Lagrange action is also banal-conformally invariant (albeit in a more complicated 
way):
$$
\widetilde{\fI} = 
\int\int_{\Sigma}\d\Sigma\{\widetilde{\fT_{\mbox{\scriptsize\tt t}}} - 
                           \widetilde{\fV}\}\d\widetilde{\vec{\tea}} 
= \int\int_{\Sigma}\d\Sigma
\left\{
\widetilde{M}_{\Gamma\Delta}\widetilde{\Star_g}Q^{\Gamma}\widetilde{\Star_g}Q^{\Delta}/2 
- \widetilde{\fV}
\right\}
\d\widetilde{\vec{\tea}} = 
\int\int_{\Sigma}\d\Sigma
\{\Omega^2M_{\Gamma\Delta}\Omega^{-2}\Star_gQ^{\Gamma}\Omega^{-2}\Star_gQ^{\Delta}/2 
- \Omega^{-2}{\fV}\}\Omega^2\d \vec{\tea}
$$
$$
= 
\int\int_{\Sigma}\d\Sigma\{M_{\Gamma\Delta}\Star_gQ^{\Gamma}\Star_g Q^{\Delta}/2  - 
{\fV}\}\d \vec{\tea} = \fI \mbox{ } .  
$$

\noindent 4) The Euler--Lagrange equations of motion following from (\ref{prod}), will clearly be 
invariant under the {\it full banal-conformal transformation} $(\fT, \fW, \Star) \longrightarrow 
(\widetilde{\fT}, \widetilde{\fW}, \widetilde{\Star}) = (\Omega^2\fT, \Omega^{-2}\fW, \Omega^{-2}\Star)$, 
as the action that they follow from is.  

\noindent
[Note that in the case of the relational formulation, the banal transformation of $\fT$, $\fW$ directly 
implies the full banal transformation, so that these are not here distinct entities.]  

\mbox{ }

\noindent
This seemingly trivial extra fact does however generate some interesting comments when one looks at the 
details of the cancellations at the level of the Euler--Lagrange equations themselves.

Let us begin with the largely-sufficient finite and trivially spatially relational case (i.e. mechanics 
with temporal relationalism only, fully reduced 1- or 2-d relational particle mechanics and  
minisuperspace).  
Then the Euler--Lagrange equations are
\beq
{D^2 Q^{\Gamma}/D\vec{\tea}}^{2} \equiv
\Star\Star Q^{\Gamma}  + \Gamma^{\Gamma}\mbox{}_{\Lambda\Sigma}\Star Q^{\Lambda}\Star Q^{\Sigma} 
= \pa^{\Gamma}\fW \mbox{ } , 
\label{*}
\eeq
which is the geodesic equation modulo the right-hand-side term ($D/D\vec{\tea}$ being the absolute 
derivative with respect to $\vec{\tea}$).  
The path of motion is not in general an affinely-parametrized geodesic [`simplification B)'], however a 
banal conformal transformation to a such exists, in the following sense. 

\noindent
Case 1: if $\fW$ is prescribed as a constant, then (\ref{*}) {\sl is} the geodesic equation, which is 
the case in mechanics if $\fV$ is constant and in (for the moment) minisuperspace GR if $\fR$ is 
constant.  
Indeed, this corresponds to having an action proportional to $\int\d\lambda\sqrt{\fT}$ and so to 
$\int\d s$ for $\d s^2$ the line-element corresponding to the kinetic metric ${M}_{\Gamma\Delta}$. 

\noindent
Case 2: if not, banal conformal-transform with $\Omega^2 = k\fW$ for $k$ constant so $\fT 
\longrightarrow \widetilde{\fT} = k\fW\fT$ and $\fW \longrightarrow \widetilde{\fW} = \fW/k\fW = 1/k$.  
This corresponds to obtaining an action $\fI \propto  = \int\d\widetilde{s}$ and passing from 
$\tea^{\sJ\sB\sB}$ to $\tea^{\sa\sf\sf-\sg\se\so}$, that is from banal conformal factor $\Omega^2 = 1$  
to banal conformal factor $\Omega^2 = k\fW$.  
Thus simplifications A) and B) are related by a banal transformation.  
Moreover, case 2 has range of validity caveats \cite{Magic, BT} for regions containing zeros of $\fW$ as 
the conformal transformation's definition precludes these; infinities and non-smoothnesses of $\fW$ can 
likewise be disruptive.
In the minisuperspace case, this paragraph's contents were spelled out by Misner \cite{Magic} following 
more partial mention in earlier work of DeWitt \cite{DeWitt70}.
In mechanics, this is in e.g. \cite{Lanczos, B94I}.

I then ask the following question. 
How does performing two transformations -- conformal transformation and non-affine parametrization -- 
each of which complicates the equations of motion, nevertheless work out to preserve them when applied 
together?

Understanding this requires looking at the alternative, longer proof of 4) at the level of the equations 
of motion themselves.  
By 
$$
{\widetilde{\Gamma}^{\Gamma}}\mbox{}_{\Delta\Lambda} = {\Gamma^{\Gamma}}_{\Delta\Lambda} + 
\{2{\delta^{\Gamma}}_{(\Delta} \pa_{\Lambda)}\Omega - M_{\Delta\Lambda}\pa^{\Gamma}\Omega\}/\Omega 
\mbox{ } ,
$$
symmetry and the definition of $\fT_{\vec{\stea}}$ in terms of velocities with respect to $\vec{\tea}$, 
$$
{\widetilde{\Gamma}^{\Gamma}}\mbox{}_{\Delta\Lambda}\Star Q^{\Delta}\Star Q^{\Lambda} = 
\Omega^{-4}{\Gamma^{\Gamma}}_{\Delta\Lambda}\Star Q^{\Delta}\Star Q^{\Lambda} + 
2\Omega^{-5}\{\pa_{\Delta}\Omega\Star Q^{\Delta}\Star Q^{\Lambda} - \fT_{\vec{\stea}}\pa^{\Gamma}\Omega\} 
\mbox{ } ,
$$
which, then, alongside using obvious product rule expressions for $\Star\{\Omega^{-2}\Star Q^{\Gamma}\}$ 
and $\pa^{\Gamma}\{\Omega^{-2}{\fW}\}$ gives 
$$
0 = \Star\Star Q^{\Gamma} + 
{\widetilde{\Gamma}^{\Gamma}}\mbox{}_{\Lambda\Delta}\Star Q^{\Delta}\Star Q^{\Lambda} -  
\widetilde{\pa}^{\Gamma}\widetilde{\fW} =  
$$
\beq
\Omega^{-4}
\mbox{\Large\{}
\Star\Star Q^{\Gamma} + {{\Gamma}^{\Gamma}}_{\Lambda\Delta}\Star Q^{\Delta}\Star Q^{\Lambda} 
- {\pa}^{\Gamma}{\fW}
\mbox{\Large\}} 
+ 2\Omega^{-5}  
\mbox{\Large\{}
\pa_{\Delta}\Omega \Star Q^{\Delta}\Star Q^{\Gamma} + \{\fW - \fT_{\vec{\stea}}\}\pa_{\Gamma}\Omega
\mbox{\Large\}} 
\mbox{ } . 
\eeq
Then the second big bracket cancels by the chain rule and conservation of energy in mechanics: 
$\fW - \fT_{\vec{\stea}} = \fE - \fV - \fT_{\vec{\stea}}$ or the Lagrangian form of the Hamiltonian 
constraint in (for 
the moment) minisuperspace GR: $\fW - \fT_{\sG\sR} = \mbox{Ric}(h) - 2\Lambda - \fT_{\sG\sR}$.

Next, analyze the above in terms of non-affine parametrization and conformal transformation subworkings. 
This reveals the second term in the second large bracket to be the result of non-affine parametrization.  
It cancels with the first term, which is one of two complicating terms from conformal transformation, 
the other one being the $\fT_{\vec{\stea}}$, which itself cancels with the banal conformal transformation's 
compensatory conformal scaling of $\fW = \fE - \fV$ by the conservation of energy or the Hamiltonian 
constraint.
This is therefore an interesting configuration space generalization of the result by which null 
geodesics conformally map to null geodesics \cite{Wald}.  
There, the first conformal complication is balanced by a change of what is the suitable affine 
parametrization, while the second one vanishes by the geodesic being null with respect to the indefinite 
spacetime metric.
In our case, the first of these cancellations continues to occur with the same interpretation, but 
what was the null combination (and thus working for indefinite metrics only) becomes, in the 
configuration space context, the kinetic term whether for indefinite or definite kinetic metrics, and 
the null condition becomes replaced by the energy or Hamiltonian constraint (granted the banal conformal 
transformation's compensatory scaling of the potential factor $\fW$).  
Thus `in indefinite spaces null geodesics conformal-map to null geodesics' becomes `in configuration 
spaces of whatever signature, paths of motion banal-conformal map to paths of motion'.

Let us now afford a slight generalization to finite models with non-trivial spatial relationalism.  
At least Euclidean and similarity relational particle mechanics then have as equations of motion
\beq
D_g^2 Q^{\Gamma}/D\vec{\tea}^2 \equiv
\Star_g \Star_g Q^{\Gamma}  + \Gamma^{\Gamma}\mbox{}_{\Lambda\Sigma}\Star_g Q^{\Lambda}\Star_g 
Q^{\Sigma} = \pa^{\Gamma}\fW \mbox{ } , 
\label{**}
\eeq
for $D_g/D\vec{\tea}$ the G-gauged absolute derivative, and then the preceding analysis carries 
through under $\Star \longrightarrow \Star_g$.  
To the extent that the previous paths of motion were geodesics, the current paths are 
`geodesics provided that we suitably align the constituent snapshots by auxiliary G-transformations'.

We have determined that solely non-affinely parametrizing or solely rescaling the kinetic metric 
complicate the equations of motion away from the simple form (\ref{*}) or (\ref{**}) that using emergent  
Jacobi--Barbour--Bertotti time places them in, while performing both of these operations alongside the 
compensating $\fW$ rescaling preserves this simple form, choice of emergent time indeed being nonunique 
up to this `banal' freedom.
Thus, if one's problem requires rescaling {\sl or} non-affinely parametrizing, one's problem may 
permit one to `complete' the required transformation to a full banal conformal transformation, whereby 
the effect of solely rescaling or solely non-affinely parametrizing kicking one out of the form
(\ref{*}) or (\ref{**}) is circumvented, and so emergent time's being a banal covector leads to a 
robustness result for its property of giving simple equations of motion.

Preservation under full banal transformation means that $\vec{\tea}$ corresponding to {\sl any} $\Omega$ 
carries out simplification A).  
One can then pick $\Omega^2 = k\fW$ so that $\widetilde{\pa}^{\Gamma}\widetilde{\fW} = 
\widetilde{\pa}^{\Gamma}k = 0$, and then $\widetilde{*} = \Omega^{-2}* = \{k\fW^{-1}\}* = 
\{k\sqrt{\fW\fT}\}^{-1}\Circ$;
%
%This recovers what is given about B) in \cite{B94I}.   
% 
i.e. so that simplification B) -- taking affine geodesic form rather than additionally 
containing a $\pa^{\Gamma}\fW$ term -- {\sl also} holds.  
Thus one has gone from physics with a restricted class of affine parameters under which the equations of 
motion take the form (\ref{*}) or (\ref{**}) to physics with a restricted class of banal conformal 
factors under which the equations of motion take geodesic form.  
Each of these, moreover, is nonunique up to a constant multiplicative time-scale (evident in the 
specification of the geodesic equation forming $\Omega$) and an additive constant time-origin (evident 
since what a power of $\Omega$ scales is $\d/\d \vec{\tea}$ and so $\vec{\tea}$ itself has 
an addititive constant of integration more freedom than $\Omega$ itself \cite{SemiclI, SemiclII}).  
These retain one's civilization's freedom of choice for calendar year zero and unit of time, as should 
be the case.

Finally, affine transformations send $\tea_{\so\sll\sd}$ to $\tea_{\sn\se\sw}(\tea_{\so\sll\sd})$ 
subject to 

\noindent
I) nonfreezing and monotonicity, so $\d \tea_{\sn\se\sw}/\d \tea_{\so\sll\sd} > 0$ which can be encoded by 
having it be a square of a quantity $\ttQ$ with no zeros in the region of use, and 

\noindent
II) this derivative and hence $\ttQ$ being a physically-reasonable function 
(to stop the transition damaging the equations of motion).  
But this can be recast as $\d/\d \tea_{\sn\se\sw} = \ttQ^{-2}\d/\d \tea_{\so\sll\sd}$, by which 
(and other 
properties matching\footnote{It 
%%%%%%%%%%%%%%%%%%%%%%%%%%%%%%%%%%%%%%%%%%%%%%%%%%%%%%%%%%%%%%%%%%%%%%%%%%%%%%%%%%%%%%%%%%%%%%%%%%%%%%%%%
may be interesting to find out whether the one restricts the other's function space more than usual.}) 
%%%%%%%%%%%%%%%%%%%%%%%%%%%%%%%%%%%%%%%%%%%%%%%%%%%%%%%%%%%%%%%%%%%%%%%%%%%%%%%%%%%%%%%%%%%%%%%%%%%%%%%%%
we are free to identify this $\ttQ$ with $\Omega$, so any affine transformation is of a form that extends 
to a (full) banal conformal transformation.  
If one then chooses to `complete' it to a full banal conformal transformation, the above calculation can 
be interpreted as the extra non-affine term being traded for a $\fT$ term by having an accompanying 
conformal transformation of the kinetic metric, and then this being traded for $\pa^{\Gamma}\fW$ by 
energy conservation and the compensating banal conformal transformation of $\fW$.
Thus the freedom to affinely-transform the geodesic equation on configuration space can be viewed 
instead as the freedom to (fully) banal-conformally transform a system's equation of motion.  
{\sl Thus the relational approach's simplicity notion for equations of motion has the same mathematical 
content as prescribing an affine rather than non-affine parameter for the geodesic equation on 
configuration space}. 
Thus the banally related $\vec{\tea}$ corresponds to `the set of (generally) nonaffine parameters 
for the geodesic-like equation of motion on configuration space', while $\tea^{\sa\sf\sf-\sg\se\so}$ 
indeed remains identified with the much more restricted set (unique up to a multiplicative constant 
time-scale and an additive constant time-origin) of affine parameters for the geodesic equation on 
configuration space.

Then part of the argument for emergent time being fixed by the universe's contents" \cite{B94I} is 
lost as it is revealed to contain an arbitrary factor.
But one can then get back that preciseness by making a choice. 
$\tea^{\sJ\sB\sB}$ ($\Omega = 1$, so that $\fE$ carries no nonconstant factors) and  
$\tea^{\sa\sf\sf-\sg\se\so}$ are then interesting such choices.

\mbox{ }

\noindent 5) In preparation for the passage to QM in Sec 3, the conjugate momenta are the 
banal-conformal invariant expressions 
\beq
\widetilde{P}_{\Delta} = 
\sqrt{  {\widetilde{\fW}}/{\widetilde{\fT}}  }\widetilde{M}_{\Gamma\Delta}\Circ_g q^{\Gamma} = 
\widetilde{M}_{\Gamma\Delta}\widetilde{\Star}_gQ^{\Gamma} = 
{M}_{\Gamma\Delta}{\Star_g}Q^{\Gamma} = P_{\Delta}
\mbox{ } .  
\label{20}
\eeq
Thus what does banal-scale concerns, a fortiori, configuration space rather than phase space.

%\mbox{ }

\noindent 6) One obtains as a primary constraint resultant from the reparametrization invariance of 
the action a quadratic constraint (\ref{set}).  
Now, as $N^{\Gamma\Delta}$ is the inverse of $M_{\Gamma\Delta}$, it scales as a banal-conformal covector 
\beq
N^{\Gamma\Delta} \longrightarrow \widetilde{N}^{\Gamma\Delta} = \Omega^{-2}N^{\Gamma\Delta} \mbox{ } .  
\label{22}
\eeq
Combining (\ref{20}) and (\ref{22}), the quadratic constraint (\ref{set}) is a banal covector.

%\mbox{ }

\noindent 7) In cases with nontrivial configurational relationalism there are also linear constraints 
from variation with respect to G-auxiliaries (Sec 1.2).  
By (\ref{20}) the linear momentum constraint of GR, ${\cal L}_{\mu}$,  and the relational particle 
mechanics constraints, $\underline{\ttP}$, $\underline{\ttL}$ and $\ttD$, are banal-conformally 
invariant.

%\mbox{ }

\noindent 8) The mechanics Hamiltonian $\fH$ is the left hand side of the quadratic energy constraint 
(\ref{set}) and as such scales as a banal covector.  
One then integrates this with respect to $\vec{t}$.  
If one looks to extend this prescription to relational particle mechanics, given that the linear 
constraints are banal-conformally invariant, one finds that one needs to either: build the total 
almost-Hamiltonian\footnote{Total Hamiltonians  
%%%%%%%%%%%%%%%%%%%%%%%%%%%%%%%%%%%%%%%%%%%%%%%%%%%%%%%%%%%%%%%%%%%%%%%%%%%%%%%%%%%%%%%%%%%%%%%%%%%%%%%%%
have constraints appended by Lagrange multiplier coordinates.
In the cyclic velocity auxiliary picture that relationalism instead implies, one has rather what I term 
an almost-Hamiltonian \cite{FEPI} with cyclic velocities in place of multipliers, which terminology 
I use since velocities still appear in it, albeit only the velocities of {\sl auxiliary} quantities.}
%%%%%%%%%%%%%%%%%%%%%%%%%%%%%%%%%%%%%%%%%%%%%%%%%%%%%%%%%%%%%%%%%%%%%%%%%%%%%%%%%%%%%%%%%%%%%%%%%%%%%%%%%  
by using $\Star\underline{a}$ and $\Star\underline{b}$ (and --$\Star{c}$ in the similarity case) as the 
appending auxiliaries to preserve homogeneity of banal transformation: 
\beq 
\fA_{\st\so\st\sa\sll} = \fH + \Star\underline{a}\cdot\underline{\ttP} + 
                               \Star\underline{b}\cdot\underline{\ttL} 
         \mbox{ } \mbox{ } ( \mbox{ } - \Star{c}         \,               \ttD \mbox{ } ) \mbox{ } .
\eeq
Or, have $\fH$ carry a `lapse' prefactor $\d \vec{T}/\d \lambda$ and then use 
$\Circ\underline{a}$ and $\Circ\underline{b}$ (and $-\Circ{c}$ in the similarity case) as the appending 
auxiliaries, producing an almost-Hamiltonian that is now banal-conformally invariant, 
\beq 
\overline{\fA}_{\st\so\st\sa\sll} = \Circ{\vec{T}}\fH + 
                                    \Circ\underline{a}\cdot\underline{\ttP} + 
                                    \Circ\underline{b}\cdot\underline{\ttL} 
            \mbox{ } \mbox{ } ( \mbox{ } - \Circ{c} \,\ttD \mbox{ } ) \mbox{ } .
\eeq
These two views are, moreover, physically equivalent, since the first is to be integrated over 
$\vec{T}$ and the second over $\lambda$.

For GR, ${\cal H}$ scales as a banal covector and the velocity of the instant, $\Circ{\mI}$ (or the lapse, $\alpha$) as a banal vector.  
Thus it is entirely straightforward to append the banal-scalar ${\cal H}_{\mu}$ using the banal-scalar 
auxiliary $\Circ{F}^{\mu}$ (or $\beta^{\mu}$), to make either the relational total GR almost-Hamiltonian 
\beq 
\fA^{\sG\sR}_{\st\so\st\sa\sll} = \int\d^3x\{\Circ{\mI}{\cal H} + {\Circ\mF}^{\mu}{\cal L}_{\mu}\} \mbox{ } ,
\eeq
or the Arnowitt--Deser--Misner total GR Hamiltonian 
\beq 
\fH^{\sG\sR}_{\st\so\st\sa\sll} = \int\d^3x\{\alpha{\cal H} +    {\beta}^{\mu}{\cal L}_{\mu}\} \mbox{ } .
\eeq
[Comparing the last two paragraps, relational particle mechanics likewise admit Hamiltonians if formulated partially 
non-relationally by use of multiplier coordinates in place of their cyclic velocities.  
Also note that GR comes already-parametrized, so there is no primed-unprimed ambiguity therein.]

%========================================================================================================
%========================================================================================================
\section{Banal-invariance in finite QM (relational mechanics, minisuperspace)}
%========================================================================================================
%=======================================================================================================

At the quantum level, those constraints which remain become wave equations.  
In this paper's models, this always includes a quadratic constraint of form (\ref{set}).  
This contains a product of $P_{\Gamma}$ and $Q^{\Gamma}$ terms, 
$N^{\Gamma\Delta}(Q^{\Lambda})P_{\Gamma}P_{\Delta}$, which picks up ordering issues in passing to QM.  
Assume that $Q^{\Gamma} \longrightarrow \hat{Q}^{\Gamma} = Q^{\Gamma}$, $P_{\Gamma} \longrightarrow 
\hat{P}_{\Gamma} = -i\hbar\pa_{\Gamma}$.

I first consider the more widely well-defined finite case, but leave the working in a general notation 
that covers both relational particle mechanics and minisuperspace.
The {\it Laplacian ordering} at the QM level of the classical 
$N^{\Gamma\Delta}(Q^{\Lambda})P_{\Gamma}P_{\Delta}$ combination is  
\beq
D^2 = \frac{1}{\sqrt{M}} \frac{\pa}{\pa {Q}^{\Gamma}}
\left\{
\sqrt{M}N^{\Gamma\Delta}\frac{\pa}{\pa{Q}^{\Delta}}
\right\} \mbox{ } . 
\eeq  
This has the desirable property of (straightforwardly) being independent of coordinate choice on the 
configuration space $\fQ$ \cite{DeWitt57}. 
This property is not, however, unique to this ordering; one can reorder to include a Ricci scalar 
curvature term so as to have $D^2 - \xi\,\mbox{Ric}(M)$ \cite{DeWitt57, Magic, HP86, CZ, Halliwell, 
Moss, Page91, RT}. 
It is then well-known \cite{Wald} that there is a unique choice of $\xi$ 
(dependent on the dimension $k \geq 2$ of the mathematical space in question\footnote{I exclude 
1-d configuration spaces as physics 
%%%%%%%%%%%%%%%%%%%%%%%%%%%%%%%%%%%%%%%%%%%%%%%%%%%%%%%%%%%%%%%%%%%%%%%%%%%%%%%%%%%%%%%%%%%%%%%%%%%%%%%
concerns changes of one configurational variable with respect to another physical variable, in 
which sense configuration space dimension $\geq$ 2 is required.
This is the case at the classical level, without it internal time makes no sense and it is also needed 
at the quantum level, at least for such as \cite{AF} the semiclassical approach and records theory 
approach to the Problem of Time in Quantum Cosmology.
}) 
%%%%%%%%%%%%%%%%%%%%%%%%%%%%%%%%%%%%%%%%%%%%%%%%%%%%%%%%%%%%%%%%%%%%%%%%%%%%%%%%%%%%%%%%%%%%%%%%%%%%%%%
 that leads to the production of a conformally-invariant operator and hence of a conformally-invariant 
operator-ordering in the quantum application \cite{Magic, HP86, CZ, Halliwell, Moss, 
Page91, RT}.
Moreover, in the present paper I identify this conformal invariance associated with operator ordering as 
being the same as the banal-conformal invariance that is simple and natural in classical relational 
product-type actions, whereby demanding this operator ordering can be seen as asking to extend this 
simple and natural invariance to hold also at the quantum level.  
The operator with the desired properties is, specifically,  
\beq
{D}_{\sC}^2 = \frac{1}{\sqrt{M}} \frac{\pa}{\pa {Q}^{\Gamma}}
\left\{
\sqrt{M}N^{\Gamma\Delta}\frac{\pa}{\pa{Q}^{\Delta}}
\right\} 
- \frac{k - 2}{4\{k - 1\}}\mbox{Ric}(M) \mbox{ }    
\eeq 
where $k$ is the dimension of $\fQ$. 
Moreover, this operator by itself is still not banal-conformally invariant: it is furthermore required 
that the wavefunction of the universe $\Psi$ that it acts upon itself transforms in general tensorially  
under \fQ-conformal transformations \cite{Wald}, 
\beq
\Psi \longrightarrow \widetilde{\Psi} = \Omega^{\frac{2 - k}{2}}\Psi \mbox{ } .
\eeq

Some simpler cases are as follows. 

\mbox{ }

\noindent
1) For models with 2-d configuration spaces such as for the minisuperspace models \cite{HP86, Page91, 
Louko, Barvin}, quantum similarity relational particle mechanics of 3 particles in the plane \cite{08II, 
+tri} or of 4 particles on the line \cite{AF}, and of 3 particles on the line with scale \cite{SemiclIII, 
scaleQM}, the conformal value of $\xi = \{k - 2\}/4\{k - 1\}$ collapses to zero, so that 
Laplacian ordering and conformally invariant wavefunctions suffice.  

\mbox{ }

\noindent 
2) For models with zero Ricci scalar, the conformal ordering coincides with the Laplacian one.
E.g. banal-conformally flat models can be arranged to have this, of which the Euclidean relational 
particle mechanics of 3 particles in the plane \cite{08I} or of N particles on a line \cite{scaleQM} 
are examples.  

\mbox{ }

\noindent
3) If a space has constant Ricci scalar, then the effect of a $\xi \mbox{Ric}(M)$ term, conformal or 
otherwise, is just something which can be absorbed into redefining the energy in the case of mechanics.
In particular, this is the case for relational particle mechanics in 1-d as their configuration spaces 
are $\mathbb{S}^{\sN - 2}$ which are clearly of constant curvature, and for relational particle 
mechanics in 2-d as their configuration spaces are $\mathbb{CP}^{\sN - 2}$ which are Einstein 
\cite{FORD} and hence of constant Ricci scalar curvature. 
Parallelly, were the Ricci scalar constant in a GR model, it could likewise be absorbed 
into redefining the cosmological constant.

\mbox{ }

\noindent
However, almost all other minisuperspace models and relational particle mechanics models (e.g. \cite{08III}) have 
configuration space dimension $\geq 3$ for which the choice of a value of $\xi$ is required.  
The present paper is written partly in support of the choice of ordering made in \cite{08II, AF, 08III} 
and more complicated relational particle mechanics (see e.g. \cite{FORD}).  

\mbox{ } 

Next, if one sends $\fH\Psi = \fE\Psi$ to $\widetilde{\fH}\widetilde{\Psi} = 
\widetilde{\fE}\widetilde{\Psi} = \{\fE/\Omega^2\}\widetilde{\Psi}$, one's eigenvalue problem has a 
weight function $\Omega^{-2}$ which then appears in the inner product: 
\beq
\int_{\Sigma}\widetilde{\Psi_1}\mbox{}^*\widetilde{\Psi_2}\Omega^{-2}\sqrt{\widetilde{M}}d^{k}x \mbox{ } .  
\label{5}
\eeq  
This inner product additionally succeeds in being banal-conformally invariant, being equal to (c.f. 
\cite{Magic} for the minisuperspace case)  
\beq
\int_{\Sigma}\Psi_{1}\mbox{}^*\Omega^{\frac{2 - k}{2}} \Psi_{2}\Omega^{\frac{2 - k}{2}}
             \Omega^{-2}\sqrt{M}\Omega^{k}\d^{k}x  = 
\int_{\Sigma}\Psi_{1}\mbox{}^*\Psi_{2}\sqrt{M}\d^{k}x \mbox{ } 
\label{FORK}
\eeq
in the banal representation that is mechanically natural in the sense that $\fE$ comes with the trivial 
weight function, 1.

\mbox{ }

Generally, $\widehat{\widetilde{\fH}} = \widetilde{\widehat{\fH}}$ is not in a simple sense self-adjoint 
with respect to $\widetilde{\langle} \mbox{ } | \mbox{ } \widetilde{\rangle}$, while the 
mechanically-natural $\widehat{\fH}$ is, in a simple sense, with respect to $\langle \mbox{ } | \mbox{ } \rangle$.  
More precisely, this is in the sense that 
\beq
\int\sqrt{M}\d^kx \Psi^*D^2\Psi = \int\sqrt{M}\d^kx \{D^2\Psi^*\}\Psi + \mbox{boundary terms },
\eeq 
which amounts to self-adjointness if the boundary terms can be arranged to be zero, whether 
by the absence of boundaries in the configuration spaces for 1 and 2 dimensional relational particle mechanics \cite{FORD} 
or by the usual kind of suitable fall-off conditions on $\Psi$.   
This is not shared by the $\Omega$-inner product as that has an extra factor of $\Omega^{-2}$, which in 
general interferes with the corresponding move by the product rule ($\sqrt{M}$ does not interfere thus 
above, since the Laplacian is built out of derivatives that are covariant with respect to the metric 
$M_{\Gamma\Delta}$.)
However, on the premise that solving $\widetilde{\fH}\widetilde{\Psi} = \widetilde{\fE}\widetilde{\Psi}$ is equivalent to 
solving $\fH\Psi = \fE\Psi$, the banal-conformal transformation might at this level be viewed as a 
sometimes-useful computational aid, with the answer then being placed in the preceding paragraph's 
banal representation for further physical interpretation.  

\mbox{ }

What about the case of theories with further, linear constraints?
In the case of relational particle mechanics, choosing conformal ordering before and after dealing with 
the linear constraints do not appear to agree in general (so that arguing for conformal ordering by 
itself is not a guarantee of unambiguously fixing an ordering).  
As I consider the structure of the configuration space to impart lucid knowledge whenever the reduction 
can be done, I would favour performing the reduction and then conformal-ordering the reduced 
configuration space Hamiltonian, as in \cite{08II, 08III, AF}. 
In the Dirac quantization approach for relational particle mechanics (`quantize then constrain'), N.B. 
that sending $\ttP\Psi = 0$ to $\widetilde{\ttP}\widetilde{\Psi}$ does cause an alteration as $\ttP$ is a 
differential operator.
The same is the case for the zero total angular momentum constraint $\ttL$ and the dilational constraint 
$\ttD$.  
 On the other hand, the reduced quantization approach (`constrain and then quantize') is free of this issue.

Barvinsky \cite{Barvin} investigated for what ordering these two approaches coincide.  
On the other hand, e.g. Ashtekar, Horowitz, Romano and Tate \cite{AHRT} have argued for inequivalence of these two 
approaches to quantization.  
In any case, to 1 loop (first order in $\hbar$) Barvinsky argues that the Laplacian ordering will do 
the trick.    
Then, as the $\xi \mbox{Ric}(M)$ term contributes only to $O(\hbar^2)$ so that conformal ordering will 
likewise do to get equivalence between Dirac and reduced quantization equivalence to 1 loop.

%========================================================================================================
%========================================================================================================
\section{Comments on quantum geometrodynamics itself}
%========================================================================================================
%========================================================================================================

Sec 3 contains the use of conformal ordering in minisuperspace, which I would argue is already an 
important and useful case on which there is substantial literature.   
For infinite theories like GR, one has not an ordinary but a {\it functional} Laplacian,   
\beq
{\cal D}^2 = \frac{1}{\sqrt{M}} \frac{\delta}{\delta {\cal Q}^{\Gamma}}
\left\{
\sqrt{M}N^{\Gamma\Delta}\frac{\delta}{\pa{\cal Q}^{\Delta}}
\right\} \mbox{ } .   
\eeq  
Using this as one's ordering for $N^{\Gamma\Delta}({\cal Q}^{\Lambda}){\cal P}_{\Gamma}{\cal P}_{\Delta}$ 
continues to have the desirable property of being independent of the coordinate choice on the 
configuration space. 
As before, this property is not, however, unique to this ordering: one can include a Ricci scalar 
curvature term so as to have ${\cal D}^2 - \xi\,\mbox{Ric}(M)$.  
Proceeding analogously to before, there is then a unique banal-conformally invariant choice among these 
orderings:   
\beq
{{\cal D}_{\sc}}^2 = \frac{1}{\sqrt{M}} \frac{\delta}{\delta {\cal Q}^{\Gamma}}
\left\{
\sqrt{M}N^{\Gamma\Delta}\frac{\delta}{\delta{\cal Q}^{\Delta}}
\right\} 
- \frac{k - 2}{4\{k - 1\}}\mbox{Ric}(M) \mbox{ } , 
\label{33}
\eeq
so long as $\Psi$ itself transforms in general tensorially under the conformal transformation 
\beq
\Psi \longrightarrow \widetilde{\Psi} = \Omega^{\frac{2 - k}{2}}\Psi \mbox{ } .
\label{34}
\eeq
There is now a snag in that $k$ is infinite so (\ref{34}) becomes ill-defined; however in the operator  
(\ref{33}) the coefficient of Ric($M$) merely tends to 1/4, while the cancellation of $k$ in 
working (\ref{FORK}) also continues to hold in this case, and it is the outcome of this (including its 
operator expectation counterpart), rather than $\Psi$ itself, that has physical meaning.  
This gives a Wheeler--DeWitt equation of the form 
\beq
\hbar^2
\left\{ 
\frac{1}{\sqrt{{\cal M}}}\frac{\delta}{\delta h_{\mu\nu}}
\left\{
\sqrt{{\cal M}}{\cal N}^{\mu\nu\rho\sigma}\frac{\delta}{\delta h_{\rho\sigma}}
\right\} 
- \frac{1}{4}\mbox{Ric}({\cal M}) 
\right\}
\Psi + \sqrt{h}\{\mbox{Ric}(h) - 2\Lambda\}\Psi = 0 \mbox{ } . 
\eeq

Also in full GR, due to the presence of the linear momentum constraint and the previous Sec's insight 
from relational particle mechanics' analogous linear constraints, conformal order before and after 
reduction may differ given the insight from the relational particle mechanics toy models.  
Moreover one cannot in general perform the reduction here, so the conformal 
order within the Dirac-type quantization scheme may be questionable.

%========================================================================================================
%========================================================================================================
\section{Conclusion}
%========================================================================================================
%========================================================================================================

\noindent
Mechanics and fundamental physics at the classical level can be considered in terms of temporally 
relational product-type actions, and doing so is useful in considering whole-universe situations -- 
the setting for quantum cosmology.  
These readily exhibit a banal-conformal invariance under compensating rescalings of the configuration 
space metric and the potential (alongside the total energy in the case of mechanics).  
The classical equations of motion resulting from product-type actions simplify for a particular form 
of emergent time.  
In mechanics, this amounts to a recovery of Newtonian time from relational premises, while in GR this 
amounts to a recovery of proper time or cosmic time in the various contexts where relevant.    
In this paper, we found that this emergent time itself scales when a banal conformal transformation 
is performed.  
Then how a more complicated manifestation of banal-conformal invariance is present in the more 
commonly used difference-type actions can be deduced, provided that the notions of time in these scale 
in the same way as the emergent time does.
I also clarified that the simplifying effects on the equations of motion through use of the emergent 
time (e.g. Jacobi--Barbour--Bertotti time) and those through use of affine parametrization of geodesics 
and dynamical trajectories are linked via a straightforward (albeit apparently hard to spot) working 
hinging on conservation of energy.

Suppose then that one chooses to retain this banal-conformal invariance -- simple and natural 
from the perspective of relational product actions at the classical level -- upon passing to the quantum level.  
Furthermore, let the theoretically-desirable and fairly standard tenet that quantum mechanics be 
independent of choice of coordinates on configuration space $\fQ$ be adhered to.  
Then these combine to pick out the operator ordering based on the conformally-invariant modification of 
the Laplacian.
While this operator ordering has been suggested by others previously (as documented in Sec 1.3), this 
is the first paper pointing out the relational underpinning for it.  
As how one operator-orders has consequences for the physical predictions of one's theory, and there is 
no established way to prescribe the operator ordering in the case of (toy models of) quantum gravity, 
this stronger motivation for conformal ordering is of wide interest.\footnote{There does 
%%%%%%%%%%%%%%%%%%%%%%%%%%%%%%%%%%%%%%%%%%%%%%%%%%%%%%%%%%%%%%%%%%%%%%%%%%%%%%%%%%%%%%%%%%%%%%%%%%%%%%%%%  
remain the caveat that QM in general and Quantum Gravity in particular may have other restrictions on 
orderings from such as well-defined existence and self-adjointness of quantum Hamiltonians and of other 
important quantum operators.    
Then possibly another such technical condition could turn out to be incompatible with conformal ordering 
at least for some theories/models, but to the best of our knowledge, to date nobody has found any such.}  
%%%%%%%%%%%%%%%%%%%%%%%%%%%%%%%%%%%%%%%%%%%%%%%%%%%%%%%%%%%%%%%%%%%%%%%%%%%%%%%%%%%%%%%%%%%%%%%%%%%%%%%%%

As regards applications to simple toy models, for 2-d configuration spaces, conformal ordering becomes 
indistinguishable from the also sometimes advocated Laplacian ordering, while the difference becomes 
minor for manifolds with constant Ricci scalar.  
Nor is there any distinction between these to 1 loop in the semiclassical approach.  
However, more complicated modelling situations \cite{08III, FORD} do have a distinction between 
Laplacian and conformal orderings. 
What is conformal ordering depends in detail (to more than 1 loop) on whether one Dirac-quantizes or 
reduced-quantizes.  
This distinction is already visible in finite but linearly-constrained relational particle models.

Inner products, which are the directly physically meaningful constructs in quantum theory, are found to 
be suitably banal-conformally invariant.  
Taking these to be primary, that the scaling of the waverfunction itself (required for the conformal 
modification of the Laplacian operator to actually form a conformally invariant combination) is 
formally infinite in cases with infinite configuration space dimensions, appears unproblematic.  
In particular, this suggests an ordering for the Wheeler-DeWitt equation of full geometrodynamics.  

\mbox{ } 

\noindent{\bf Acknowledgments} I thank Visiting Prof Julian Barbour, Dr Harvey Brown and Ms Anne 
Franzen for discussions. 

\vspace{3in}

%=====================================================BIBLIOGRAPHY==========================================================================

\end{document}